\documentstyle[12pt]{article}
\newcommand{\draft}{
        \renewcommand{\baselinestretch}{1.0}%
        \small\normalsize%
}

\draft
\begin{document}
\title{\bf Birth rates of different types of neutron star and possible 
evolutions of these objects}
\author{Oktay H. Guseinov$\sp{1,2}$
\thanks{e-mail:huseyin@gursey.gov.tr},
A\c{s}k\i n Ankay$\sp1$
\thanks{e-mail:askin@gursey.gov.tr}, \\
Sevin\c{c} O. Tagieva$\sp3$
\thanks{email:physic@lan.ab.az}
\\ \\
{$\sp1$T\"{U}B\.{I}TAK Feza G\"{u}rsey Institute} \\
{81220 \c{C}engelk\"{o}y, \.{I}stanbul, Turkey} \\
{$\sp2$Akdeniz University, Physics Department,} \\ 
{Antalya, Turkey} \\
{$\sp3$Academy of Science, Physics Institute,} \\
{Baku 370143, Azerbaijan Republic} \\
}

\date{}
\maketitle
\begin{abstract}
\noindent
We estimate the spatial densities of different types of neutron star near 
the Sun. It is shown that the distances of dim isolated thermal neutron 
stars must be on average about 300-400 pc. The combined birth rate of 
these sources together with radio pulsars and dim radio quiet neutron 
stars can be a little more than the supernova rate as some of the dim 
isolated thermal neutron stars can be formed from dim radio quiet neutron 
stars and radio pulsars. Some of these sources must have 
relations with anomalous X-ray pulsars and soft gamma repeaters. In order 
to understand the locations of different types of neutron star on the 
P-\.{P} diagram it is also necessary to take into account the differences 
in the masses and the structures of neutron stars. 
\end{abstract}

\section{Introduction}
Existence of single neutron stars with different physical properties is 
very well known. But the amount of data about different types of neutron 
star except radio pulsars is small. Actually, the change in the beaming 
factor together with the value of and the change in the angle between the 
magnetic field and the rotation axes are not well known. It is also not 
clear how the relation between the characteristic time ($\tau$) and the 
real age changes in time. Therefore, the birth rate, initial periods and 
values of the real magnetic field are not known well. So, it is difficult 
to understand what the evolutionary tracks of pulsars on the P-\.{P} 
diagram must be and where these tracks end. As there are not many 
available observational data for the other types of neutron star and since 
they show some exotic phenomena such as $\gamma$-ray bursts, it is 
difficult to understand their nature. Below, we analyse birth rates of 
different types of neutron star, their locations on the P-\.{P} diagram 
and possible evolutions of these objects.   

\section{Analysis of the data about the birth rates of DITNSs and  
supernova rate around the Sun}
According to Yakovlev et al. (2002) and Kaminker et al. (2002) neutron 
stars cool down to T=4.6$\times$10$^5$ K (or kT=40 eV) in t$\le$10$^6$ yr. 
According to Haberl (2003, and references therein), there are 7 X-ray dim 
isolated thermal neutron 
stars (DITNSs) located within 120 pc around the Sun and only one DITNS, 
namely RX J1836.2+5925, is located at a distance of 400 pc. All of these 
8 radio-quiet objects have T$>$4.6$\times$10$^5$ K. On the other 
hand, there are only 5 radio pulsars with characteristic times  
smaller than 10$^7$ yr in the cylindrical volume around the 
Sun with a radius of 400 pc and height 2$\mid$z$\mid$=400 pc, where 
$\mid$z$\mid$ is the distance from the Galactic plane. Among these 5 
pulsars, Vela has $\tau$$\sim$10$^4$ yr and the other 4 pulsars have 
3$\times$10$^6$$<$$\tau$$<$5$\times$10$^6$ yr. The real ages of 1-2 
of these 4 radio pulsars may be smaller than 10$^6$ yr. We have included 
Geminga pulsar (B0633+1748) in Table 1 to compare it with different types 
of neutron star, since this pulsar has properties in between the  
properties of DITNSs and of young radio pulsars. 

In order to make a reliable estimation of the number density of the 
sources in each considered volume, it is necessary to consider how these 
objects move in the Galaxy. Practically, for pulsars with such ages the 
space velocity must not decrease in the gravitational field of the Galaxy. 
In order to demonstrate this in a simple and reliable way, we can use the 
observational data about the scale height and the average peculiar  
velocity of old stars near the Sun. 

The average values of 
$\mid$z$\mid$ and $\mid$V$_z$$\mid$ for population II stars which belong 
to the old disk, the intermediate and the halo are, respectively, 400 pc 
and 16 km/s, 700 pc and 25 km/s, and 2000 pc and 75 km/s (Allen 1991). 
Ages of these stars are close to 10$^{10}$ years. They are in dynamical 
equilibrium and their ages are large enough to make many oscillations in 
the gravitational field of the Galaxy. 

Let us consider, in the first approximation, their oscillations in a 
homogeneous field in the direction perpendicular to the Galactic plane. In 
this case
\begin{equation}
V_z = A \sqrt{z} \hspace{0.2cm} cm/sec
\end{equation}   
and the period of oscillation is 
\begin{equation}
T = \frac{8V_z}{A^2} \hspace{0.15cm} sec
\end{equation}
where A is a constant which is related to the gravitational field 
intensity near the Sun. If we use the average values of $\mid$z$\mid$ and 
$\mid$V$_z$$\mid$ for different subgroups of population II stars as 
written above, then we find values of A very close to each other showing 
the reliability of the approximation. Therefore, we 
can adopt a value of A=5.2$\times$10$^{-5}$ cm$^{1/2}$s$^{-1}$, which is 
the average value for the 3 subclasses of population II objects, for all 
objects located up to about 2000 pc away from the Galactic plane and we 
can also use this value for pulsars. If we put the average value 
of $\mid$V$_z$$\mid$=170-200 km/s for pulsars in eqn.(1), then we find 
T=(1.2-1.7)$\times$10$^9$ yr. As we see the age values of  
young radio pulsars are very small compared to the average 
period of pulsar oscillation. So, we can adopt that pulsars with ages 
10$^6$-10$^7$ yr move practically with constant velocity which they gain 
at birth. 

Since the progenitors of neutron stars have scale height of about 60 pc, 
most of the neutron stars are born close to the Galactic plane. 
Therefore, they can go 200 pc away from the plane in about 10$^6$ yr. As 
the Sun is located very close to the Galactic plane, considering the 
value of z of the Sun in the analysis does not change the results and so 
can be neglected. 

There are 23 supernova remnants (SNRs) within about 3.2 kpc around 
the Sun with surface brightness $>$10$^{-21}$ Wm$^{-2}$Hz$^{-1}$sr$^{-1}$ 
and the ages of these SNRs are not greater than (3-4)$\times$10$^4$ yr 
(Green 2001; Ankay et al. 2003). So, the 
number of neutron stars which were born in the last 10$^6$ yr within the 
region of 400 pc around the Sun must be about 11. Only 2 or 3 of 
these 11 neutron stars can go more than 200 pc away from the Galactic 
plane. Therefore, the considered region may contain about 8-9 neutron 
stars. But the birth rate of radio pulsars is up to 3-4 times 
smaller than the supernova explosion rate in our galaxy (Ankay et al. 
2003). So, we can expect only 3 radio pulsars with an age up to 
10$^6$ yr located in this region and this is in accordance with the data 
mentioned above.

According to Haberl (2003) (see also the references therein), 
surface temperatures of DITNSs are 40-95 eV and their ages 
must not be more than 10$^6$ yr (Kaminker et al. 2002; Yakovlev et al. 
2002) as they are in the cooling stage. Guseinov et al. (2003a) 
include most of these sources (for which there exist 
considerably more information) in the list of isolated radio quiet neutron 
stars which were observed in X-ray band. We have included in Table 1 
only a few very important data taken from the tables given in Guseinov 
et al. (2003a) together with some new data. In some cases, the distance 
values of DITNSs are about 2-3 times larger than the ones given in 
Haberl (2003) and in some other papers (see also the references in 
Guseinov et al. 2003a and Haberl 2003). Is it acceptable to adopt such 
smaller values of distance as given by Haberl (2003) for these objects? 

The statistics about the SNRs and the radio pulsars in the considered 
region are very poor, but the number density of DITNSs for the case 
of small distances (Haberl 2003) is very large so that there exists a 
contradiction. If the data about the distances and ages of DITNSs were 
reliable, then their birth rate would turn out to be about 4 times more 
than the supernova explosion rate. 

Since the measured space velocities of all types of neutron star are very 
large compared to the space velocities of O and B-type stars, there is no 
doubt about the origin of neutron star formation which is due to the 
collapse of the progenitor star together with supernova explosion. For 
example, RX J0720.4-3125 at a distance of 200 pc (which may actually be 
$\sim$300 pc) has a tangential velocity of about 100 km/s calculated from 
the proper motion $\mu$=97$\pm$12 mas/yr (Motch et al. 2003).
PSR J0538+2817 has proper motion $\mu$=67 mas/yr and has a transverse 
velocity in the interval 255-645 km/s at a distance of 1.2 kpc (Kramer et 
al. 2003). Therefore, the birth rate of neutron stars can not exceed the 
rate of supernova explosion. How can we explain the contradiction between 
the neutron star birth rate and the supernova explosion rate? Note that 
Haberl (2003) does not assume all the DITNSs to be only cooling neutron 
stars but also discusses other possibilities, for example the accretion 
from interstellar gas. 
But as a rule, today the cooling origin of the X-ray radiation is 
considered and this is reliable (Haberl 2003). In order to solve the 
problem about the difference in the birth rates it is necessary to adopt 
either larger distance values or longer lifetimes for DITNSs. 

\section{The distances of DITNSs} 
Since the theory of 
cooling of neutron stars has been well developed and all the DITNSs have 
T$>$4.6$\times$10$^5$ K, we can say that these neutron stars have ages 
less than 10$^6$ yr (Pavlov et al. 2002a,b; Kaplan et al. 2003a,b; Pavlov 
\& Zavlin 2003; Yakovlev et al. 2003). As the ages of these objects 
are small, their number density must be small. Therefore, it is better to 
adopt larger values of distance for these neutron stars. 

Before beginning to discuss how to adopt distance 
values, it is necessary to analyse the data of DITNSs. These 
data vary a lot from one observation to another. The most recent data 
about DITNSs are given in Haberl (2003) and Guseinov et al. (2003a).   

For all DITNSs there exist data about their temperatures and it is known 
that they have approximately blackbody radiation (Pavlov et al. 2002b; 
Haberl 2003). In this work, we have used the temperature values given in 
Pavlov et al. (2002b) and Haberl (2003) which are more in accordance with 
the other data that they are more reliable. 

The cooling curves of neutron stars, which are obtained from the fit of 
the data of different pulsars on the surface temperature versus age diagram, 
are given in various articles (see for example Yakovlev et al. 2002). 
Since the general form of the cooling curves given by different authors 
is approximately the same in all the works, we will use the data given in 
Yakovlev et al. (2002). 

DITNS RX J1605.3+3249 (Motch et al. 1999) is located in a region of sky 
(l=53$^o$, b=48$^o$) where it is easier to 
determine the temperature with small uncertainty. There are 2 different 
distance estimations for this source, 0.1 kpc (Haberl 2003) and 0.3 kpc 
(Kaplan et al. 2003b). 
We may adopt reliable distance values for any source which has a 
spectrum close to the blackbody using luminosity values. The temperatures 
of DITNS RX J1605.3+3249 and of radio pulsars J1057-5226 and J0659+1414 
which have similar soft X-ray spectrums are 0.092, 
0.070 and $\le$0.092 keV, respectively (Yakovlev et al. 2003). Therefore,   
the luminosity of RX J1605.3+3249 must not be smaller than the luminosity 
of these radio pulsars (Becker \& Aschenbach 2002; Guseinov et al. 2003b), 
i.e. it must not be less than about 10$^{33}$ erg/s. 
The luminosity of RX J1605.3+3249 at 0.1 kpc is 1.1$\times$10$^{31}$ erg/s 
(Haberl 2003), so that, its luminosity at 0.3 kpc must be about 
10$^{32}$ erg/s. On the other hand, Geminga pulsar (B0633+17), which has 
temperature of about 0.045 keV, has L$_x$=1.05$\times$10$^{31}$ erg/s in 
0.1-2.4 keV band (Becker \& Trumper 1997). This also suggests to adopt a 
larger L$_x$ value for RX J1605.3+3249. So, we can adopt a distance value 
of 0.3 kpc (Kaplan et al. 2003b) or even a larger value which is more 
reliable. 

Following the same path, we can adopt distance values for other DITNSs; 
the distances of J0420.0-5022, J0720.4-3125, J0806.4-4123, J1308.8+2127, 
and J214303.7+065419 must be about 3-4 times larger than the distance 
values given in Haberl (2003). All the new distance values and other 
reliable data are represented in Table 1.   

\section{Where must DITNSs be located on the P-\.{P} diagram?} 
There are 6 single radio pulsars (J1952+3252, J0659+1414, J0117+5914, 
J1057-5226, J0358+5413 and J0538+2817) with characteristic times in the 
range 10$^5$-6$\times$10$^5$ yr 
from which X-ray radiation has been observed (Becker \& Aschenbach 2002; 
Possenti et al. 2002). These pulsars are located up to 2 kpc from the 
Sun. The luminosities of these radio pulsars are in the range 
10$^{31}$-5$\times$10$^{33}$ erg/s (Guseinov et al. 2003b). 
There are also 3 radio pulsars 
(J0826+2637, J0953+0755 and J1932+1059) with characteristic times 
3$\times$10$^6$$<$$\tau$$<$2$\times$10$^7$ yr (it is necessary to 
remember that value of $\tau$ may considerably exceed the real age value). 
The distances of these 3 radio 
pulsars are less than 400 pc and their luminosities are $<$10$^{30}$ erg/s 
(Possenti et al. 2002). 

As seen from Table 2, the X-ray luminosities of all the 8 DITNSs and 
Geminga are in the interval 10$^{30}$-1.7$\times$10$^{32}$
erg/s. Among these sources, Geminga has the lowest luminosity 
(T$_{eff}$ value of Geminga is also very small, see Table 1) and the 
$\tau$ value of Geminga is 3.5$\times$10$^5$ yr. Taking these data into 
consideration, the ages of DITNSs can be adopted as 10$^5$-10$^6$ yr in 
accordance with the age values calculated from the cooling 
models. Moreover, these sources are nearby objects and there is not any 
pulsar wind nebula around them nor any SNR to which they are connected; 
this also shows that the ages of these objects must be greater than 10$^5$ 
yr. On the other hand, pulsar wind nebula is present around the 
neutron stars with the rate of rotational energy loss 
\.{E}$>$5$\times$10$^{35}$ erg/s and with L$_x$(2-10  
keV)$>$5$\times$10$^{32}$ erg/s (Guseinov et al. 2003b). 
Naturally, it may be possible to observe 
pulsars with smaller values of \.{E} and L$_x$ (2-10 keV) which are 
located closer to the Sun. Taking these facts into account, we can assume 
that \.{E} values of DITNSs must be less than about 3$\times$10$^{35}$ 
erg/s (constant \.{E}=3$\times$10$^{35}$ erg/s line is shown on the 
P-\.{P} diagram, see Figure 1). 

It is well known that the difference between $\tau$ and the real age 
can be significant for very young pulsars (Lyne \& Graham-Smith 1998). On 
the other hand, for single-born old pulsars $\tau$ must be approximately 
equal to the real age if the evolution takes place under the condition 
B=constant. But none of the DITNSs is connected to a SNR, so that, DITNSs 
must be located in the belt between $\tau$=3$\times$10$^4$ yr and 
$\tau$=10$^6$ yr lines on the P-\.{P} diagram if the condition B=constant 
is satisfied. 

For 5 of the 9 DITNSs (including Geminga) represented in Table 1, the spin 
periods (P) have been measured. Four of these objects have spin periods 
greater than 8 s and Geminga pulsar has P=0.237 s. As known, the period 
values of anomalous X-ray pulsars (AXPs) and soft gamma repeaters (SGRs) 
are greater than 5 s (see for example Mereghetti 2001; Guseinov 
et al. 2003a and the references therein). The small value of 
\.{P}=5.4$\times$10$^{-13}$ s/s belongs to AXP 1E2259+586 which has 
P=6.98 s. Therefore, all single neutron stars with P$>$10 s must be 
related to AXPs and SGRs. The P=10 s line is displayed in Figure 1 to 
show the two separate regions in which DITNSs can be located. The DITNSs 
with $\tau$$<$10$^6$ yr may have values of P$>$10 s (see Table 1) and this 
is possible if the values of \.{P} are very large. Therefore, these 
objects may be the evolutionary continuations of SGRs and AXPs. 
Naturally, their birth rates should be in agreement with each other for 
this assumption to be true. 

The locations of pulsars Geminga, RX J0720.4-3125 and RX J1308.8+2127 
on the P-\.{P} diagram are shown in Figure 1. From the 
position of Geminga pulsar it is seen that this pulsar evolves similar to 
the radio pulsars with B=10$^{12}$-10$^{13}$ G. The position of pulsar RX 
J0720.4-3125 is not within our chosen interval of $\tau$=10$^4$-10$^6$ yr, 
but as this pulsar has a high value of kT$_{eff}$ (see Table 1) its 
real age (according to the cooling models) must be smaller than its $\tau$ 
value. So, there may be magnetic field decay or some other reason for 
this pulsar (i.e. n$>$3, where n is the braking index). 
Pulsar RX J1308.8+2127 is located in the SGR/AXP region on the P-\.{P} 
diagram so that this pulsar seems to have a relation with the SGR/AXP 
class of neutron stars. Pulsars RX J0806.4-4123 and RX J0420.0-5022 with 
ages $<$10$^6$ yr have large values of P. Although their \.{P} values are 
not known, they must be located on the upper part of the P-\.{P} diagram 
and their positions must not be lower than the position of J0720.4-3125, 
if the P values are correct.  

\section{Birth rates of different types of neutron star near the Sun}
In section 2, we have mentioned that the birth rate for all the types of 
neutron star, in other words the supernova rate, must be about 11 in 
10$^6$ yr in the region up to 400 pc from the Sun. In the same region, 
the birth rate of radio pulsars can be about 3-4 in 10$^6$ yr. 

In Figure 1, we have plotted all the 9 radio pulsars with 
$\tau$$\le$4$\times$10$^4$ yr which are connected to SNRs and located at 
distances up to 3.5 kpc (Guseinov et al. 2003c). There are also 2 other 
radio pulsars, J1048-5832 and J1837-0604, in this region with such values 
of $\tau$ but without any connection with SNRs (Guseinov et al. 2003b). 
From these data it follows that the birth rate 
of radio pulsars is 3.6 in 10$^6$ yr in the region up to 400 pc. It is 
necessary to take into consideration that for such young radio pulsars the 
beaming factor is close to 1 and the influence of the luminosity function 
on the estimations is small. Note that the searches of pulsars near the 
Sun in the central regions of SNR shells and the searches of pulsar wind 
nebulae are 
considerably better than the searches of pulsars under the surveys. Also 
note that in some cases pulsars have been found after observing point 
X-ray sources in the central parts of SNRs. Therefore, we can adopt that 
the birth rate in the region up to 400 pc from the Sun is about 3-4 radio 
pulsars in 10$^6$ yr. 

It is necessary to take into account that in the region with distance up 
to 3.5 kpc from the Sun, there are also 6 dim radio quiet neutron stars 
(DRQNSs) which are connected to SNRs (Table 1). The locations of 2 
of these objects, 1E1207.4-5209 and RXJ0002+6246, are shown in Figure 1. 
These objects have considerably large values 
of $\tau$ compared to the ages of the SNRs in which they are located, so 
that, they have different evolutionary tracks compared to 
other radio pulsars which are connected to SNRs. If we assume that 
all DRQNSs are also radio pulsars with 10$^{12}$$<$B$<$10$^{14}$ G, then 
they must have low radio luminosities and/or the direction of their 
radio radiation does not pass through the line of sight. On the other 
hand, all of them may have large P and $\tau$ values and significantly 
different evolutionary tracks compared to other pulsars with the same 
magnetic field, because there exist some other important differences 
between DRQNSs and most of the radio pulsars. Practically, all the radio 
pulsars which are connected to SNRs and which have similar ages as DRQNSs 
have pulsar wind nebula (Guseinov et al. 2003b). None of the DRQNSs has 
such property. Therefore, these DRQNSs have \.{E}$<$3$\times$10$^{35}$ 
erg/s. The positions of 1E1207.4-5209 and RX J0002+6246 on the P-\.{P} 
diagram require magnetic field decay or some additional ideas for the 
pulsar models (in Fig.1 the location of RX J0002+6246 has been found 
using the condition $\tau$ = age of the SNR). Therefore, the number of 
SNRs which contain ordinary and other types of pulsars with similar 
properties for d$\le$3.5 kpc is 17. By this approach, we give the upper 
limit for pulsar birth rate in the region up to 400 pc which is 5.5 in 
10$^6$ yr.  

In section 3, we have adopted up to 3-4 times larger distance values for 
DITNSs and this gives us the possibility to estimate the birth rate of 
this type of neutron star as 9 in 10$^6$ yr in the same region with a 
radius of 400 pc. These sources have P$>$0.1 s, \.{E}$<$3$\times$10$^{35}$ 
erg/s and, according to the cooling theories, ages between 
3$\times$10$^4$-10$^6$ yr that they must be in 
later stages of the evolution of single neutron stars with initial 
magnetic field B$>$10$^{12}$ G (see the locations of these sources in 
Figure 1). But how can we explain such a large birth 
rate for these sources which is comparable with the supernova rate? 
First, note that the statistical data are poor. Second, the actual ages of 
the SNRs may be not up to 4$\times$10$^4$ yr but up to 3$\times$10$^4$ yr 
(see Table 1). On the other hand, the rate of supernova must be a little  
more if we take into account the SNRs which have low surface brightness 
during their evolution. In this case, the birth rate of DITNSs must 
roughly be equal to the birth rate of radio pulsars and DRQNSs together.   

As seen from Figure 1, there exist 5 single radio pulsars which have been 
detected in X-ray band and have 10$^5$$<$$\tau$$<$6$\times$10$^5$ yr and 
5$\times$10$^{11}$$<$B$<$1.1$\times$10$^{12}$ G. They are located in the 
region up to 2 kpc from the Sun and 2 of them, J0659+1414 and J0538+2817, 
are most probably connected to S type SNRs (Kramer et al. 2003; Guseinov 
et al. 2003b and the references therein). If we also consider that 2 
such pulsars may go far away from the Galactic plane that they can be 
missed in the surveys, the total number of the pulsars in the considered 
volume turns out to be 7. Therefore, the birth rate of this type of 
pulsar in the region with d$\le$400 pc is not more than 0.6 in 10$^6$ yr. 

From the estimations done in this section, we see that approximately 60\% 
of the SNRs with surface brightness $>$10$^{-21}$  
Wm$^{-2}$Hz$^{-1}$sr$^{-1}$ are connected to normal pulsars and DRQNSs. 
The rate of birth for DITNSs is also approximately equal to 60\% of the 
rate of supernova explosion. Therefore, the neutron stars with ages 
approximately $<$5$\times$10$^5$ yr, which show themselves as radio pulsar 
or DRQNS in SNRs, may mainly transform to DITNSs.   

The numbers of radio pulsars with effective values of 
magnetic field B$\ge$10$^{13}$ G and B$\ge$3$\times$10$^{12}$ G which 
have $\tau$$<$10$^6$ yr and d$\le$3.5 kpc are 5 and 32, respectively (ATNF 
pulsar catalogue 2003; Guseinov et al. 2002). From these data, the birth 
rates of radio pulsars with B$\ge$10$^{13}$ G and  
B$\ge$3$\times$10$^{12}$ G located up to 400 pc from the Sun must be  
considerably more than 0.06 and 0.4, respectively, in 10$^6$ yr, because 
the $\tau$ values may be several times larger than the real ages.   

In the region up to 8 kpc from the Sun, there are 4 AXPs and one SGR 
and the ages of these objects must not be larger than 5$\times$10$^4$ yr 
(Mereghetti 2001; Guseinov et al. 2003a). Therefore, the 
birth rate of AXPs and SGRs in the considered cylindrical volume with 
the radius of 400 pc in 10$^6$ yr is not less than 0.15. This is about 60 
times smaller than the supernova rate in the same volume and the time 
interval, but it approximately coincides with the birth rate of radio 
pulsars with effective value of B$\ge$10$^{13}$ G. As seen from Figure 1, 
only 3 pulsars, J1740-3015, J1918+1444 and J1913+0446 (in the 
order of increasing value of P), are located in the volume with a radius 
of 3.5 kpc and with $\tau$$\le$10$^5$ yr.     

\section{Discussion and Conclusions}
It is known that, different types of neutron star which have different 
properties are born as a result of supernova explosion: radio pulsars, 
DRQNSs, DITNSs, AXPs and SGRs. There exist PWN and often SNR shell around 
very young single radio and X-ray pulsars which have 
\.{E}$>$5$\times$10$^{35}$ erg/s (Guseinov et al 2003b). 
There may exist only the shell around pulsars with  
\.{E}$<$3$\times$10$^{35}$ erg/s, DRQNSs and AXPs the ages of which are 
less than 10$^5$ yr. No shell nor PWN has been found around some 
of the very young pulsars (e.g. J1702-4310 and J1048-5832) and SGRs. 
Absence of the shell or PWN around DITNSs must be considered normal 
because they have considerably large ages. 

In section 4, we have adopted $\sim$3-4 times larger distance values, 
compared to the values usually adopted (Haberl 2003), for most of the 
DITNSs  
and this gives the possibility to decrease their birth rate down to the 
sum of the birth rates of radio pulsars and DRQNSs. Birth rate of these 3 
types of neutron star together is approximately equal to the rate of 
supernova. The combined birth rate of these 3 types of neutron star may be 
more than the supernova rate, because some of the DITNSs may be formed as 
a result of the evolution of DRQNSs and radio pulsars. Birth rate of AXPs 
and SGRs, which belong to the same class of objects, is about 60 times 
smaller than the supernova rate and it is about same as the birth rate of 
radio pulsars with effective values of B$\ge$10$^{13}$ G. 

As seen from Table 1, the period values of 4 DITNSs are very large, though 
their ages are smaller than 10$^6$ yr. From this situation, there arises 
a possibility of a relation between some of the DITNSs and AXPs/SGRs. 
Naturally, we must take into account the position of each DITNS on the 
P-\.{P} diagram to show the relation between some of these objects and 
AXPs/SGRs. 

Existence of radio pulsars with n$<$3 and with real ages smaller 
than $\tau$ (for young pulsars) show that the condition B=constant is not 
satisfactory in all cases and this is well seen in Figure 1. Most of the 
pulsars with 10$^5$$<$$\tau$$<$10$^7$ yr are not in the belt 
B=10$^{12}$-10$^{13}$ G where most pulsars are born in; often, there 
occurs magnetic field decay. But the evolutions of AXPs/SGRs and some 
of the DITNSs according to the field decay approach (the magnetar 
model) lead to bimodality in the number of neutron stars versus the 
magnetic field distribution. On the 
other hand, the time scale of the magnetic field decay must be very short. 
This shows that the large effective B values of these objects and the 
shape of their evolutionary tracks must be related mainly to the masses of 
and the density distributions in the neutron stars and also to the 
activity of the neutron star. This is also necessary 
to understand the different positions of radio pulsars which are 
connected to SNRs and of DRQNSs on the P-\.{P} diagram despite the fact 
that they have similar ages. If the evolution under the condition 
B=constant were true, then they would be located along $\tau$=constant 
belt, but not along the constant magnetic field belt. 

We think that there is a possibility to get rid of these difficulties and 
to understand the large X-ray luminosities and also the bursts of 
AXPs/SGRs. It is necessary to assume the birth of neutron stars with 
masses 
about half of the maximum mass values found from the given equations of 
state and rotational moment. In principle, it must be easy to identify 
such smaller mass neutron stars as they are far away from hydrodynamical 
equilibrium. They must have an elipsoidal shape due to rotation and 
possibly they do not rotate as a rigid body. In this case, the young 
pulsar may especially demonstrate itself when the angle between the 
magnetic field and the rotation axes is close to 90$^o$. 
In such a case, a considerably larger effective value of magnetic field 
can be produced as compared to the real magnetic field.


\draft

\begin{flushleft}
\begin{tabular}{cccccccc}
\multicolumn{8}{l}{\bf Table 1 - The data of DRQNSs and DITNSs. Ages of 
the} \\ 
\multicolumn{8}{l}{\bf SNRs connected to these objects are given in the 
4$^{th}$} \\ 
\multicolumn{8}{l}{\bf column. References are shown at the end of the 
table.} \\ 
\multicolumn{8}{l}{\bf Some data adopted in this work are given 
without reference.} \\ \hline 
Names & P & $\tau$ & t & d & kT$_{eff}$ & L$_{\hbox{\footnotesize 
x}}$10$^{32}$ & L$_{\hbox{\footnotesize x}}$/\.{E} \\
& s & kyr & kyr & kpc & keV & erg/s & \\ \hline
1E1207.4-5209 & 0.424 & 340- & 7-20 & 1.8 & 0.11 & 10d$_2^2$ & $\sim$0.1 
\\
G296.5+10.0 & [4] & 480 & [5,15] & 2 & [1] & (0.5-6) & \\
S & & [4,5] & & [6] & 0.25 & [7] & \\ 
& & & & 2.1 & (0.5-6) & & \\ 
& & & & [7] & [7] & & \\ \hline
1E0820-4247 & & & 3-4 & 2 & 0.15 & 12 & \\
RXJ0822-4300 & & & [5,11,12] & [8] & [1] & (0.1-2.4) & \\
Puppis A & & & & & 0.44 ? & [8,9] & \\
G260.4-3.4 & & & & & [10] & & \\
S & & & & & & & \\ \hline
CXOJ2323+5848 & & & 0.32 & 3.2 & & & \\
Cas A & & & [25] & [2] & & & \\
G111.7-2.1 & & & & & & & \\ \hline 
RXJ0002+6246 & 0.2418 & & 10-20 & 3.5 & 0.10 & 2 & 0.0002- \\
$\gamma$-ray source & [11] & & [5,7] & [11] & [1] & (0.5-2) & 0.0006 \\
G117.7+0.6 & & & & [11,2] & \\ \hline
RXJ0007.0+7302 & & & 10-24 & 1.4 & & 0.15 & \\
G119.5+10.2 & & & [5,8-10] & [2] & & (0.1-2.4) & \\
CTA 1 & & & & & & [2] & \\ \hline
RXJ2020.2+4026 & & & 6-10 & 1.5 & & 9 & 0.0009- \\
$\gamma$-ray source & & & [4-6] & [12] & & [12] & 0.00009 \\
G78.2+2.1 & & & & & \\
$\gamma$ Cyg & & & & & \\ \hline
CXOJ0852-4615 & & & 0.7-2 & 1 & 0.40 & 2.3 & \\
G266.2-1.2 & & & [13,14] & [13] & [13] & (0.4-6) & \\ 
& & & & & & d=1kpc & \\ 
& & & & & & [24] & \\ \hline
Geminga & 0.237 & 350 & & 0.16 & 0.048 & 0.01 & 0.00003 \\
B0633+1748 & [21] & [9] & & & [1] & (0.6-5) & \\ 
& & & & & & [23] & \\ \hline

\end{tabular}
\end{flushleft}

\clearpage
\begin{flushleft}
\begin{tabular}{cccccccc}
\multicolumn{8}{l}{\bf Table 1 (continued)} \\ \hline
Names & P & $\tau$ & t & d & kT$_{eff}$ & L$_{\hbox{\footnotesize
x}}$10$^{32}$ & L$_{\hbox{\footnotesize x}}$/\.{E} \\
& s & kyr & kyr & kpc & keV & erg/s & \\ \hline
RXJ1836.2+5925 & & & & 0.43 & 0.043 & 0.054 & \\
& & & & & [3] & [3] & \\ \hline
RXJ1856.5-3754 & & & & 0.13 & 0.062 & 0.17 & \\
& & & & & [3] & & \\ \hline      
RXJ0720.4-3125 & 8.39 & 3000 & & 0.3 & 0.085 & 1.7 & 55 \\ 
& [14] & [14] & & & [3] & 0.9d$_{0.3}^2$ & \\ 
& & & & & & [15] & \\ \hline
RXJ0420.0-5022 & 22.7 ? & & & 0.4 & 0.057 & 0.43 & \\ 
& [22] & & & & [3] & & \\ \hline
RX J0806.4-4123 & 11.37 ? & & & 0.4 & 0.095 & 0.91 & \\ 
& [16,17] & & & & [3] & & \\ \hline
RX J1308.8+2127 & 10.3 & 7-14 & & 0.4 & 0.091 & 0.82 & $\sim$0.2-0.6 \\
RBS 1223 & [18] & [16] & & 0.67 & [3] & & \\ 
& & & & [19] & & & \\ \hline
1RXSJ214303.7 & & & & 0.3- & 0.09 & 1-1.7 & \\
+065419 & & & & 0.4 & [3] & & \\
RBS 1774 & & & & & & & \\ \hline
RXJ1605.3+3249 & & & & 0.3- & 0.092 & 1-1.7 & \\ 
& & & & 0.4 & [3] & & \\ 
& & & & 0.3 & & & \\
& & & & [20] & & & \\ \hline
\multicolumn{8}{l}{[1] Yakovlev et al. 2002; [2] Guseinov et al. 2003c; 
[3] Haberl 2003;} \\
\multicolumn{8}{l}{[4] Bignami et al. 2003; [5] De Luca et al. 2002; [6] 
Vasisht et al. 1997;} \\
\multicolumn{8}{l}{[7] Pavlov et al. 2002a; [8] Petre et al. 1996; [9] 
Brazier \& Johnston 1999;} \\ 
\multicolumn{8}{l}{[10] Pavlov et al. 2002b; [11] Hailey \& Craig 1995; 
[12] Brazier et al. 1996;} \\ 
\multicolumn{8}{l}{[13] Kargaltsev et al. 2002; [14] Zane et al. 2002; 
[15] Kaplan et al. 2003a;} \\ 
\multicolumn{8}{l}{[16] Hambaryan et al. 2002; [17] Haberl \& Zavlin 2002; 
[18] Haberl et al. 2003;} \\ 
\multicolumn{8}{l}{[19] Kaplan et al. 2002; [20] Kaplan et al. 2003b; 
[21] McLaughlin et al. 1999;} \\
\multicolumn{8}{l}{[22] Haberl et al. 1999; [23] Halpern \& Wang 1997; 
[24] Pavlov et al. 2001;} \\
\multicolumn{8}{l}{[25] Guseinov et al. 2003d.}

\end{tabular}
\end{flushleft}

\clearpage
{\bf Figure Caption} \\
{\bf Figure 1:} Period versus period derivative diagram for different 
types of pulsar. The '+' signs denote the radio pulsars with d$\le$3.5 
kpc which are connected to SNRs. The 'X' signs show 
the positions of the radio pulsars with d$\le$3.5 kpc and 
10$^5$$<$$\tau$$<$2$\times$10$^7$ yr which have been detected in X-rays. 
The locations of 3 radio pulsars which have d$\le$3.5 kpc and 
$\tau$$<$10$^5$ yr are shown with 'circles' to make a comparison between
the birth rates (see text). DITNSs are 
represented with 'stars' and DRQNSs are displayed with 'empty squares'. 
The 'filled squares' show the positions of all AXPs/SGR in the Galaxy. 
Names of DITNSs, DRQNSs, 2 of the AXPs, and 
some of the radio pulsars are written. Constant lines of B = 
10$^{11-15}$ G, $\tau$ = 10$^{3-9}$ yr, and \.{E} = 10$^{29}$, 10$^{32}$, 
10$^{35}$, 3$\times$10$^{35}$ and 10$^{38}$ erg/s are shown. P=10 s line 
is also included (see text). 

\clearpage  
\begin{figure}[t]
\vspace{3cm}
\includegraphics{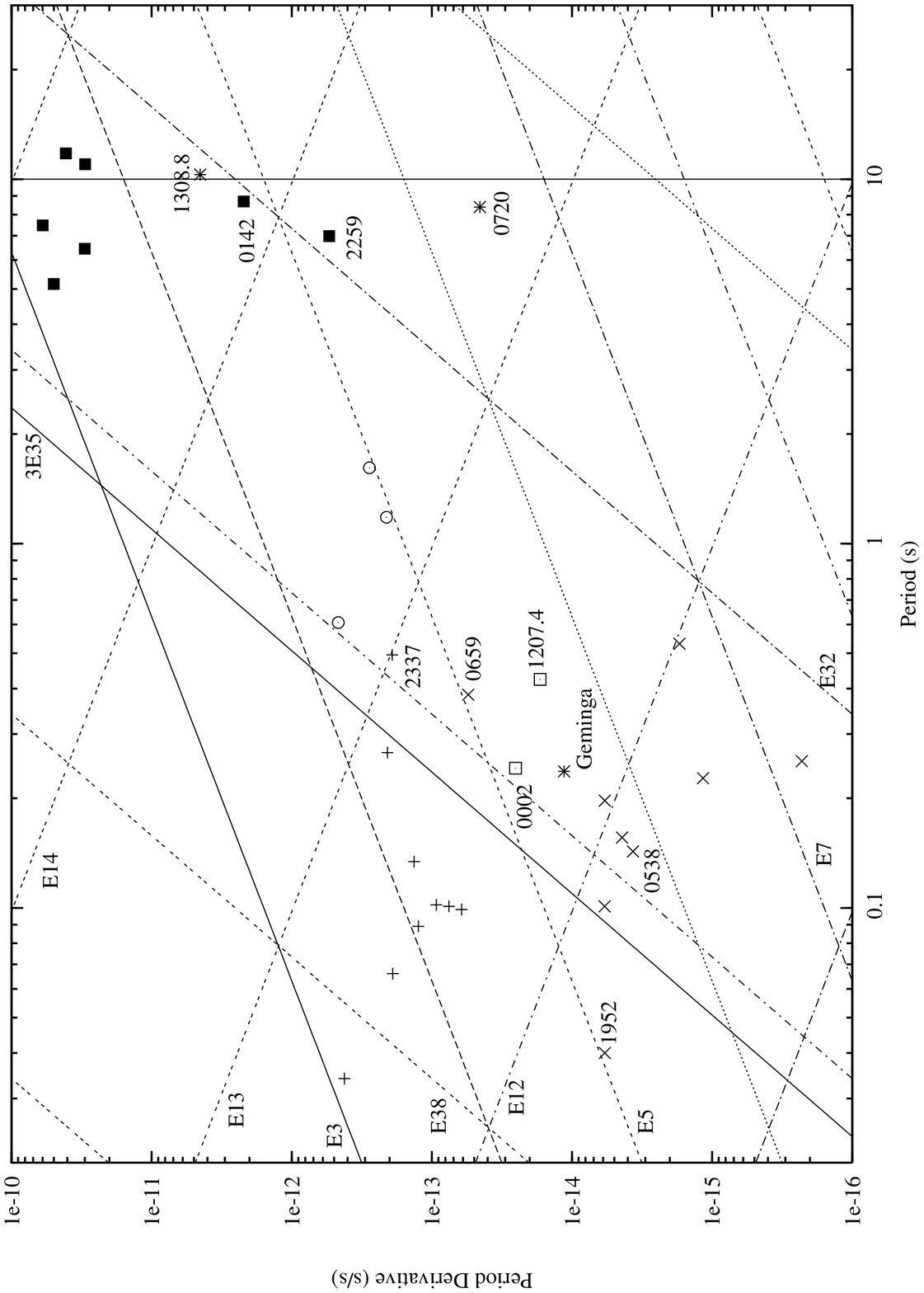}
\end{figure}

\end{document}